\documentstyle[preprint,prb,aps]{revtex}
\begin{document}
\draft
\title{Localization and conductance fluctuations in the integer quantum Hall
effect: Real--space renormalization group approach}

\author{A. G. Galstyan and M.~E.~Raikh}
\address{Department of Physics, University of Utah, 
Salt Lake City, Utah  84112}

\maketitle
\begin{abstract}
We consider the network model of the integer quantum Hall effect
transition. By generalizing the real--space renormalization group
procedure for the classical percolation to the case of quantum
percolation, we derive a closed
 renormalization group (RG) equation for the 
universal distribution of conductance of the quantum Hall
 sample at the transition. 
We find an approximate  solution of the RG equation and use it
to calculate the critical exponent of the localization length
 and the central  moments of the 
conductance distribution. The results obtained are compared with
the results of recent numerical simulations.   

\end{abstract}

\narrowtext

\section{INTRODUCTION}
The fact that at zero temperature the quantum Hall transitions are 
infinitely narrow complicates
strongly their theoretical description even without electron--electron
interactions.
By now, in theory, the quantitaive information about the localization 
of two--dimensional electronic states in
a  magnetic field was  obtained exclusively from numerical simulations. 
These simulations are carried out in one of two limits:

{\em i}) Short--range disorder   

The results obtained before 1995 are reviewed in Refs.\onlinecite{haj94} and
\onlinecite{huc95}. Different approaches employed: the 
recursive Green's function  method\cite{huc90}, evolution of the number
of states with non--zero Chern numbers with the sample size\cite{huo92}, and
 finite--size scaling analysis of the
Thouless number\cite{han95,sor96}  give consistent results for the critical
exponent, $\nu$, of the localization length.  
Direct calculations of the conductance
at the transition\cite{huo93,gam94} show that it is close to $0.5e^2/h$.
Recent developments include the study of the role of spin--orbit 
scattering\cite{han95} for spin--degenerate electrons,
localization in double\cite{sor96}-- and multilayer systems\cite{ots93},
statistics of energy levels at the transition\cite{avi95} and fractal 
properties of eigenstates at higher Landau levels\cite{ter96}.

{\em ii}) Smooth disorder

All the simulations\cite{cha88,lee93,lee94,lee'94,kag95,kag97,wan94,ruz95,cha95} were performed in the frame of the network 
model introduced  by Chalker and Coddington\cite{cha88}.
The questions addressed were essentially the same as in the case 
of the short--range disorder.
In addition calculating  $\nu$,\cite{cha88,lee93} the network
 model was used
to investigate the spin--degenerate case\cite{lee94,lee'94}, 
and the effect of Landau level mixing on the positions of 
delocalized states\cite{kag95,kag97}. It was also used to
demonstrate that this mixing does not change the universality class
of the transition\cite{wan94}, to relate 
 the longitudinal and transverse 
conductivities in the transition region\cite{ruz95} and to trace
the smearing of a single delocalized state into a metallic band in a
superlattice\cite{cha95}.

Recent works\cite{lee''94,zir94,kim96,kon96,gru96,lee96,lud94,ho96}
explored the correspondence between the network model and 
certain limits of the models that were already studied (spin chains\cite{lee''94,zir94,kim96,kon96,gru96}, Hubbard chains\cite{lee96} and Dirac fermions\cite{lud94,ho96}).

In the recent experimental paper by Cobden and Kogan\cite{cob96}
 the observation of
strong and reproducible conductance  fluctuations of a mesoscopic
sample right at the transition was reported. This observation was
also accounted for in the frame of the network model in Refs.
\onlinecite{wan96} and \onlinecite{cho96}. The result of simulations\cite{cho96} closely resemble the histogram of the conductance fluctuations in\cite{cob96}. In Ref. the calculated central moments of the conductance distribution were
extrapolated to large sample sizes, where it becomes universal.

The network model\cite{cha88} is illustrated in Fig. 1.
If a disorder is smooth, classical electron drifts along equipotentials,
which are closed and disconnected except for the percolation threshold.
For simplicity it was assumed that at any energy
equipotentials are arranged periodically.
Quantum mechanics is introduced  by allowing a strong tunnel coupling between
the orbits when they come close enough ($\sim$ or closer than the magnetic
length). 
This destroys the classical localization and makes important the interference 
of different directed paths. 
Randomness is introduced by assuming the magnetic phases, acquired in course
of drift along equipotentials, to be random. 

If a smooth potential is symmetric with respect to zero energy,
then the classical percolation threshold is also at $\varepsilon =0$.
Although this is obvious, we would like to derive the same conclusion 
following a different line
of arguments. 

The centers of closed drift trajectories  in 
Fig. 1 form a square lattice.
Classical electron switches to another trajectory, centered at a
neighboring lattice site, if its energy, $\varepsilon$, exceeds the height, 
$V_0$, of a saddle--point
separating these sites. One can say that in this case a  {\em bond} between two
sites {\em connects} them. In the opposite case,
$\varepsilon <V_0$,
the  electron  retains its orbit after passing the saddle--point.
Then one can say that the corresponding bond is removed.  
 Thus we arrive at the bond percolation problem on the square lattice. 

Suppose now, that the electron energy, which
was initially at the lowest minimum,  gradually increases. As it happens, 
more and more bonds switch on. Percolation occurs when some critical
portion of bonds connect. It is an exact result of the percolation
 theory\cite{sta92}
that for the square lattice this portion is $1/2$, which corresponds to zero
energy.

The fact that the classical picture can be reformulated as 
a percolation problem on a lattice suggests the following possibilty. 
It is known that the critical exponent for classical
percolation can be found with  high accuracy from a simple
 renormalization group (RG) calculation. 
In course of this calculation  only a small cluster is considered. 
If a similar approach could be devised for the quantum percolation,  
it would provide an approximate description 
 of the quantum Hall transition in the closed
form.  In contrast to the classical renormalization scheme, such an 
approach should deal not with ``connecting'' or ``removed'' bonds,
but characterize each bond by some scattering matrix. It should also
allow for  interference of different directed paths within a cluster.
This program is carried out in the present paper. 
We utilize a real--space RG scheme for   classical percolation, 
 proposed by Reynolds, Stanley, and  Klein\cite{rey77}, 
 and generalize it to the  quantum percolation.
This generalization is described in Sec. II. 
We derive the RG transformation, which governs the evolution
of the distribution of conductance  right at the transition,
with increasing the sample size. In Sec. III the approximate numerical
solution for the fixed point of the RG transformation
 is obtained.
This solution determines  the universal conductance distribution at 
the transition. 
 It is used for calculation of   
the exponent $\nu$ in Sec. IV. In this section we also discuss the
finite size corrections.
In Sec. V our results for the  distribution of conductance of a large sample
at the transition and in its vicinity are compared with recent numerical
simulations. 
Concluding remarks are made in Sec. VI.
\section{RENORMALIZATION PROCEDURE}
\label{II}
Let us briefly remind the renormalization procedure for the classical
bond percolation problem on the square lattice\cite{rey77,rey80,ber78}.
 In this procedure a
fragment of original lattice containing 8 bonds (Fig. 2a) is replaced by a
fragment  consisting of  two {\em superbonds}. The probability, $\tilde p$,
that a superbond {\em does connect} two {\em supersites} is expressed
 through the
corresponding probability, $p$, for original lattice in a following way.
Firstly, the supersites are connected if all five bonds, which constitute the
superbond (see Fig. 2a), connect. The probability of this is $p^5$. Secondly,
 removing
one of five bonds, does not affect the connectivity of the superbond, the
total probability of such configurations being $5p^4(1-p)$. Adding the 
probabilities for the superbond to connect when two and three original bonds
are removed, one gets the following expression for renormalized 
probability.
\begin{equation}
\label{til}
\tilde p=p^5+5p^4(1-p)+8p^3(1-p)^2+2p^2(1-p)^3.
\end{equation}
The fixed point of the transformation (\ref{til}), which determines the
percolation threshold, is $\tilde p=p=1/2$ and it coincides with the exact result. 
The equation  for the critical exponent, $\nu_c$, of the correlation length, 
$\xi_c$, emerges as a condition that $\xi_c$ is the same for original
 and for
renormalized lattices. If the lattice constant  of the original 
lattice is $a$, then $\xi_c=a(\frac{1}{2}-p)^{-\nu_c}$.  On the other hand,
the lattice constant of the renormalized lattice is $2a$, so that 
$\xi_c=2a(\frac{1}{2}-\tilde p)^{-\nu_c}$. Assuming that $(\frac{1}{2}-p)$
 and 
$(\frac{1}{2}-\tilde p)$ are small, the two relations yield 
$\nu=\biggl[\ln2/\ln\bigl(d\tilde p/dp\bigr)\biggr]_{p=1/2}$.
 Using Eq. (\ref{til}), one gets $\nu_c\approx 1.428$, which
is only 8 percent bigger than the value $4/3$, which is presumed to be exact.

As it was mentioned in  Introduction, we are interested in the 
situation when
each bond represents a saddle--point of the random potential.
Then we come to Fig. 2b,
which  can be viewed as a transformation of five original saddle--points
into a {\em super--saddle--point}.
 In
classical mechanics, $p$ is the probability  that the energy of
an electron exceeds the height of the saddle--point. In quantum mechanics,
a saddle--point is characterized by  reflection coefficient, $r$,
which is the amplitude for an incoming wave to retain 
the equipotential, or by transmission 
coefficient $t$--the amplitude to switch the equipotential.  Obviously,
$r^2+t^2=1$.  In the network,  there are two incoming and two outgoing waves
at each  saddle--point $i$. The $\bf S$--matrix, relating the
 amplitudes of outgoing waves 
to the amplitudes of incoming waves, has the form
\begin{eqnarray}\label{matr}
{\bf S}_i =
\left(\begin{array}{cc}
t_i & r_i\\
r_i & -t_i
\end{array}\right).
\end{eqnarray}
With five saddle--points we have five matrix equations. Solving the system,
we get the following expression for the effective transmission coefficient of
the super--saddle--point  
\begin{equation}
\label{t}
\tilde t=\frac{t_1t_5(r_2r_3r_4e^{i\Phi_2}-1)+t_2t_4e^{i(\Phi_3+\Phi_4)}(r_1r_3r_5e^{-i\Phi_1}-1)+t_3(t_2t_5e^{i\Phi_3}+t_1t_4e^{i\Phi_4})}{(r_3-r_2r_4e^{i\Phi_2})(r_3-r_1r_5e^{i\Phi_1})+(t_3-t_4t_5e^{i\Phi_4})(t_3-t_1t_2e^{i\Phi_3})},
\end{equation}
where $\Phi_j$ are the random phases acquired by an electron after traversing
one of four loops shown in Fig. 2b. 
This expression can be viewed as a generalization  of Eq. (\ref{til}) to the
 case of  quantum percolation.  
It appears very convenient to parametrize  $t_i$, $r_i$ as follows
\begin{equation}
\label{par}
t_i=\frac{1}{(e^{z_i}+1)^{1/2}},~~~ 
r_i=\frac{1}{(e^{-z_i}+1)^{1/2}}.
\end{equation}
Then the parameter $z_i$ can be related to the height of the saddle--point,
$V_i$, as\cite{fer87} 
\begin{equation}
\label{Gamma}
z_i=\frac{\varepsilon -V_i}{\Gamma},~~~\Gamma =\frac{|V_{xx}^{i}V_{yy}^{i}|^{1/2}l^2}{2\pi}.
\end{equation}
Here $\varepsilon$ is the energy of the electron, $V_{xx}^{i}, V_{yy}^{i}$
are the second derivatives of the saddle--point potential, and $l$ is the
magnetic length. Thus, for $\varepsilon = 0$, $z_i$ is the dimensionless height
of the saddle--point and Eq. (\ref{t}) determines the dimensionless height,
$\bf Z$$\{z_i,\Phi_j\}$, of the 
super--saddle--point for a given set of 
$z_i$ and $\Phi_j$. We now introduce the distribution function
\begin{equation}
\label{K}
K(z,\{z_i\})=\langle \delta (z-{\bf Z}\{z_i,\Phi_j\})\rangle_{\{\Phi_j\}}
\end{equation}
of the heights of super--saddle--points at given heights of original
saddle--points. The averaging in (\ref{K}) is carried out over the random
phases $\Phi_j$.\cite{sha95} The crucial test for applicability of the renormalization 
procedure to the quantum percolation is that the position of the 
delocalized state should remain unchanged after renormalization. Suppose 
that in original lattice all $z_i$ were zero. Then the delocalized state
is at $\varepsilon =0$.
 After the first step of renormalization the 
distribution of the heights of super--saddle--points is given by
$K(z,\{z_i=0\})$. This function is plotted in Fig. 3a. We see that it
is symmetric with respect to $z=0$, which justifies the further
 renormalization. Note that $z_i=0$ at the first step, implies that the
power transmission coefficient, $t_i^2$, is  equal to $1/2$ for all 
saddle--points. 
As it follows from Fig. 3a, after the first step we already 
get a substantial spread 
 in $\tilde t^2$ (the characteristic range is 0.2$\div$0.8). This is the
 result of interference of different paths, which is described by
Eq. (\ref{t}).

 Now we can write down the renormalization group equation.
Denote with $Q_n(z)$ the normalized distribution function of the heights
of super--saddle--points after the  $n$-th step of renormalization.
Then the coresponding distribution function after the step $n+1$ is
given by      
\begin{equation} 
\label{Q}
Q_{n+1}(z)=\hat T\{Q_n\}=
\int\biggl(\prod_{i=1,..,5}dz_i\biggr)~ K(z,\{z_i\})\prod_{i=1,..,5}Q_n(z_i).
\end{equation}
The fixed point, $Q(z)$, of the transformation (\ref{Q}) determines 
the distribution of conductance of a large sample at the plateau 
transition. The equation for the fixed point, $Q(z)=\hat T\{Q\}$, is a
quite complicated integral equation. In the next section we obtain its
approximate solution.

\section{APPROXIMATE SOLUTION of the RG EQUATION}

The crucial observation which allows to obtain an approximate solution for the
fixed--point distribution $Q(z)$ is that within the relevant region of change 
of the variables of integration $z_i$, which appears to be $\sum z_i^2 < 10$,
the shape of the function
$K(z, \{z_i\})$ changes rather weakly.
At the same time the position of the symmetry axis of $K(z, \{z_i\})$ depends
strongly on the values of $z_i$. In other words, the kernel of the 
transformation  (\ref{Q}) can be presented as $K(z-f\{z_i\})$, where
the function $K(z)$ is calculated for all $z_i=0$ and shown in Fig. 3.
By plotting $K(z, \{z_i\})$ for many sets $\{z_i\}$ (some examples are
shown in Fig. 3b) we had established an approximate form of the function $f$:
\begin{eqnarray}
\label{fu}
f\{z_i\}=c_1(z_1+z_2+z_4+z_5)+c_3z_3-d(z_1-z_4)(z_2-z_5)-\nonumber\\
-\lambda_1(z_1^3+z_2^3+z_4^3+z_5^3)-\lambda_3z_3^3+\lambda_2z_3(z_1^2+z_2^2+z_4^2+z_5^2),
\end{eqnarray} 
with coefficients $c_1\approx 0.416$, $c_3\approx 0.164$, 
$d\approx 0.067$, $\lambda_1\approx 0.009$,
$\lambda_2\approx 0.009$ and $\lambda_3\approx 0.003$. The quadratic term in the expansion
(\ref{fu}) is non--physical. It results from the deficiency of the
 renormalization scheme, which is the absence of complete $\pi/2$ rotational
symmetry (see Fig. 2b).
 Fortunately, the corresponding coefficient, $d$, is small enough,
so that this term has a small effect on $Q(z)$. We have verifyed it by
taking this term into account as a perturbation. This term also does not
affect the value of the critical exponent, $\nu$, since it does not change
upon a shift of all $z_i$ by the same value (see below). We have also
established (by first neglecting and then taking into account perturbatively)
that the cubic term with coefficient $\lambda_2$ leads  to a
relatively small correction to $Q(z)$. On the contrary, the cubic term
with coefficient  $\lambda_1$ appears to be very important. 

After neglecting the quadratic and the last cubic terms in (\ref{fu}) the
function $f\{z_i\}$ decomposes into the sum of functions of different 
arguments. Owing to this, a great simplification can be achieved by taking the
Fourier transform of both parts in Eq. (\ref{Q}). One gets
\begin{equation}
\label{trans}
Q_{n+1}(\omega)=
K(\omega)\biggl[\int dz_1 Q_n(z_1)e^{i\omega(c_1z_1-\lambda_1z_1^3)}
\biggr]^4 \int dz_3 Q_n(z_3)e^{i\omega(c_3z_3-\lambda_3z_3^3)},
\end{equation}  
where $Q_n(\omega)$ is the Fourier transform of $Q_n(z)$ and $K(\omega)$
is the Fourier transform of $K(z)$. Now the solution for the fixed point
$Q_n(z)=Q_{n+1}(z)=Q(z)$ can be easily obtained numerically. The result is
shown in Fig. 4. Except for the tails at $|z| >5$ it can be
well approximated by a gaussian $\sqrt{\frac{\alpha}{\pi}}exp(-\alpha z^2)$
with $\alpha \approx 0.1$. 
In the right--hand side of Eq. (\ref{Q}) there is a product of five functions
$Q(z_i)$. Having $\alpha \approx 0.1$, the  5--dimensional volume, which provides the major
contribution to the integral in (\ref{Q}), can be estimated as 
$\sum z_i^2 <10$. 
 We see that the fixed--point distribution is  broad. The 
half--width is $z\approx \pm 2.5$, which corresponds to the range
 $0.08 \div 0.92$ for the power transmission coefficient. In the 
 next section  we use the  solution  obtained for calculation of the critical
exponent $\nu$.
\section{CRITICAL EXPONENT}
\label{III}
We first address the critical exponent of the localization length.
Suppose that  before the renormalization the 
distribution of $z$ is already given by  $Q(z)$. Then for 
electron with a  small but non--zero energy, $\varepsilon$, 
this distribution is given by $Q(z-z_0)$, where $z_0=\varepsilon/\Gamma \ll 1$ 
(see Eq. (\ref{Gamma})). Substituting $Q(z-z_0)$ into the right--hand side
of the transformation (\ref{Q}) one gets a renormalized distribution,
$Q(z-\tau z_0)$, where $\tau$ is  some number, independent of $z_0$.
Upon repeating the procedure, the shift will grow as $z_0\tau^n$ and,
eventually, become $\sim 1$. At this step an electron will be strongly
localized in a renormalized lattice of super--saddle-points (within $\sim 1$
unit cell). Thus, one should identify the localization length, $\xi$, with
a lattice constant after these $n$ steps, which is equal to $2^na$.
Then the condition $z_0\tau^n \sim 1$
can be rewritten as $\frac{\varepsilon}{\Gamma}\bigl(\frac{\xi}{a}\bigr)^{(\ln \tau/\ln 2)} \sim 1$, and  one has 
$\xi \sim a\bigl(\frac{\varepsilon}{\Gamma})^{-\nu}$ with
\begin{equation}
\label{tau}
\nu =\frac{\ln2}{\ln \tau}
\end{equation} 
The remaining task is to calculate parameter $\tau$ using Eq. (\ref{Q}).
For small $z_0$ the result of substituting $Q(z-z_0)$ into the right--hand side
can be presented as
\begin{equation}
\label{sub}
Q(z)+z_0\int\biggl(\prod_{i=1..5}dz_i\biggr)\biggl(\sum_{i}\frac{\partial K}
{\partial z_i}\biggr)\prod_{i=1..5}Q(z_i)
\end{equation}
At small $z$ the first term behaves as $Q(0)+\frac{z^2}{2}Q_{zz}(0)$,
while the second term is proportional to $z$ (due to symmetry). As a result,
the maximum of the sum (\ref{sub}) will get shifted from $z=0$ to
$z=z_0\tau$, where
\begin{equation}
\label{rat}
\tau=-\frac{1}{Q_{zz}(0)}\int\biggl(\prod_{i=1..5}dz_i\biggr)\biggl(\sum_{i}
\frac{\partial^2 K}
{\partial z \partial z_i}\mid_{z=0}\biggr)\prod_{i=1..5}Q(z_i).
\end{equation}
To proceed further we adopt the same simplification, 
$K(z,\{z_i\})=K(z-f\{z_i\})$, as was adopted to find the fixed point $Q(z)$.
Then one has
\begin{equation}
\label{der}
\sum_{i}\frac{\partial^2K}{\partial z \partial z_i}\mid_{z=0}=
\frac{\partial^2 K}{\partial z^2}\bigl(-f\{z_i\}\bigr)\biggl[4c_1+c_3
-3\lambda_1(z_1^2+z_2^2+z_4^2+z_5^2)-3\lambda_3z_3^2\biggr].
\end{equation}
Note that non--physical quadratic term in (\ref{fu}) did not contribute to
(\ref{der}). After substituting (\ref{der}) into (\ref{rat}) it is again 
convenient perform the Fourier transformation and then to make use of
Eq. (\ref{trans}). Finally we get
\begin{equation}
\label{fin}
\tau=4c_1+c_3-3~\frac{\int d\omega \omega^2 Q(\omega)
\bigl[4\lambda_1 I_1(\omega)+\lambda_3 I_3(\omega)\bigr]}{\int d \omega 
\omega^2 Q(\omega)},
\end{equation}
where
\begin{equation}
\label{I}
I_{1,3}(\omega)=\frac{\int dz z^2 Q(z)e^{i\omega (c_{1,3}z-\lambda_{1,3}z^3)}}
{\int dz Q(z)e^{i\omega (c_{1,3}z-\lambda_{1,3}z^3)}}.
\end{equation}
Evaluating numerically the integrals in (\ref{I}) and then in (\ref{fin})
with $Q(z)$ found  in the previous section, we got for $\tau$ the value $\tau =1.336$, which leads to the  exponent $\nu =2.39$. 
Since $\lambda_1,\lambda_3$ were determined with certain error bars,
we repeated the calculations and for $\lambda_1= 0.01$
 (the effect of $\lambda_3$ is small)
and got $\tau = 1.306$,  $\nu =2.59$. The results of most numerical simulations
are grouped around $\nu = 2.4$.
    
Despite the agreement obtained, we cannot estimate
the accuracy of the result, since  the calculation was based
on the assumption that the shape of the kernel of the transformation (\ref{Q})
is independent of ${z_i}$. In fact the width of the distribution 
$K(z,\{z_i\})$ decreases slowly with increasing $\mid z_i\mid$, 
and, correspondingly,
$K(\omega)$ broadens. We found that at the boundary of the relevant region
 of the change of
$z_i$, $\sum z_i^2 < 10$, the width of $K(\omega)$ increases by $\sim 30$
percent.
We could not include this effect in our calculation.
To obtain the limiting estimate we repeated the calculation  assuming that
the width of $K(\omega)$ is  $30$ percent bigger for {\em all} $z_i$. This
caused the reduction of $\nu$ to $1.95$.   

In conclusion of the section let us discuss the question  how the distribution
$Q_n(z)$ approaches the fixed--point distribution $Q(z)$ with increasing
the sample size. To answer this question let us introduce the deviation 
$\delta Q_n(z)=Q_n(z)-Q(z)$ and linearize the transformation (\ref{Q}).
After using the above simplification for $K(z,\{z_i\})$ and performing the
Fourier transform, we get  
\begin{equation}
\label{lin}
\delta Q_{n+1}(\omega)=\hat F\{\delta Q_n\}=K(\omega)J_1^3(\omega)
\int dz\Biggl[4J_3(\omega)e^{i\omega(c_1z-\lambda_1z^3)} + J_1(\omega)e^{i\omega(c_3z-\lambda_3z^3)}\Biggr]\delta Q_n(z),
\end{equation}
where
\begin{equation}
\label{J}
J_{1,3}(\omega)=\int dzQ(z)e^{i\omega(c_{1,3}z-\lambda_{1,3}z^3)}. 
\end{equation}
If initial perturbation was $\delta Q_0(z)$, then after
$n$ steps it will evolve to $\hat F^n\{\delta Q_0(z)\}$.  
The relevant perturbations are the even functions of $z$, satisfying
the condition $\int dz Q_0(z)=0$ $ $  
(i. e. $Q_0(\omega \rightarrow 0)=0$), to keep $Q_n(z)$ normalized.
We were able to make the  complete analysis of the transformation  
(\ref{lin}) only in the case $\lambda_1=\lambda_3=0$.
The answer is as follows. The fluctuations for which, at 
$\omega \rightarrow 0$, $\delta Q_0(\omega)$ behaves
 as $\omega^{2m}$, $m$ being integer, decay
with increasing $n$ as $(4c_1^{2m}+c_3^{2m})^n \propto L^{-\gamma_m}$,
where $\gamma_m=\ln(4c_1^{2m}+c_3^{2m})^{-1}/\ln2$. The slowest 
decay is characterized by the  exponent  $\gamma_1 \approx 0.5$.

With finite $\lambda_1,\lambda_3$ there is no such a
simple classification.  We could only see that the exponent increases
significantly up to the value $\sim 2$.  At large $L$ this exponent
would determine the finite--size correction to the conductance
distribution and, thus, to the average conductance, variance and
all the moments. 
\section{CONDUCTANCE FLUCTUATIONS}
\label{IV}
The fixed--point  distribution $Q(z)$ is directly related to the 
distribution function of the conductance at the plateau transition.
After an appropriate number of the renormalization steps the entire
sample reduces to a single super--saddle--point; the power
 transmission coefficient of this super--saddle--point would determine 
the conductance of the sample, $\frac{e^2}{h}G$, with $G=(e^z+1)^{-1}$.
Then the distribution function of $G$ can be found from the relation
$P(G)dG = Q(z)dz$ yielding
\begin{equation}
\label{cond}
P(G)=\frac{1}{G(1-G)}Q\biggl(\frac{1}{G}-1\biggr).
\end{equation}
The distribution function $P(G)$ is shown in Fig. 5. 
It is very broad. This fact was first pointed out in the
experimental paper\cite{cob96} and then reproduced in the numerical
simulations of Cho and Fisher\cite{cho96}. As a distinctive feature,
the distribution function (\ref{cond}) has a shallow minimum at $G=1/2$.
Using this function we calculated the variance 
$\delta G = \sqrt{\langle G^2\rangle - \frac{1}{4}}$
and obtained the value $\delta G =0.33$, which agrees with the result of
Ref.\onlinecite{cho96}. To compare our results with the recent simulations
by  Wang,  Jovanovi\'{c} and D.--H. Lee\cite{wan96} we have calculated the moments of the distribution $P(G)$, defined as $A_n=\langle \bigl(G-\frac{1}{2}\bigr)^n\rangle$, where $n$ is an even integer which we have changed from
$n=2$ up to $n=20$. The results are shown in Fig. 6 together with the
values $A_2, A_4, A_6$ and $A_8$, obtained in the simulations of 
Ref.\onlinecite{wan96}. The agreement is quite good.

Finally we have studied how the variance, $\delta G(\varepsilon)$, falls off
as the energy, $\varepsilon$, deviates from the position of the delocalized
state $\varepsilon = 0$. As it was mentioned above, for non--zero
energy the distribution of $z$ is given by $Q(z-\frac{\varepsilon}{E_0})$,
where $E_0=\Gamma\biggl(\frac{L}{a}\biggr)^{-\nu}$, $L$ being the sample size.
Then one has
\begin{equation}
\label{vari}
\biggl(\delta G(\varepsilon)\biggr)^2=\int dz \frac{1}{(e^z+1)^2}
Q(z-\frac{\varepsilon}{E_0})-\Biggl( \int dz \frac{1}{e^z+1}
Q(z-\frac{\varepsilon}{E_0})\Biggr)^2.
\end{equation}
The dependence $\delta G(\varepsilon)$ is shown in Fig. 7. To compare 
it with the simulations of Ref.\onlinecite{cho96} we had to establish
the correspondence  between our dimensionless energy $\varepsilon/E_0$
and the one adopted in Ref. \onlinecite{cho96}. We did it by calculating
the average  $\langle G(\varepsilon)\rangle$ and fitting it to the curve
in\cite{cho96}. The agreement is again quite reasonable, especially
taking into account some asymmetry of the numerical results. However,
strictly speaking, the distribution of $z$ at finite $\varepsilon$
has the form $Q(z-\frac{\varepsilon}{E_0})$ only if $\varepsilon /E_0$
is smaller than $1$. 
  
Note in conclusion of the section, that  identifying the conductance
$G$ with  the transmission through the last super--saddle--point 
implies that sweeping 
the electron energy  through $\varepsilon =0$ causes an increase of $G$
from $0$ to $1$. 
This was the case for the experimental geometry of Ref. \onlinecite{cob96}. 
\section{CONCLUSION}
\label{V}
Certainly, the renormalization scheme for quantum percolation,
developed  in the present paper, provides
only an approximate description of the plateau transition. As in classical
percolation, when, in the course of renormalization, one ascribes a 
certain probability to a superbond with no care about the condition of
surrounding bonds, our quantum generalization leaves out many interference 
processes. This happens at the stage when we average over the phases, $\Phi_j$,
in Eq. (\ref{K}). By doing so we assume the phases of transmission 
coefficients of super--saddle--points to be uncorrelated, which  in reality is
not the case.
 However, to the best of our knowledge, the above
 approach is the first one in which the characteristics of the quantum Hall
transition are obtained from the solution of a certain closed equation.
By taking into account only basic interference processes at {\em each} scale,
we were able to reproduce the results of numerical simulations in which
{\em all} interference processes were taken into account. 
Our approach does not require extrapolation to infinite sizes. 
This might be important
in the cases when the result of such an extrapolation is  rather ambigous.
For example, when there are two (several) 
delocalized states {\em very close}
 in energy (Zeeman split levels, higher Landau levels).
 In this case the network 
should carry more than one channel and allow for their
 mixing\cite{lee94,lee'94,wan94,kag95,kag97}. Formal generalization of
the RG equation (\ref{Q}) to this case is straightforward, however
its approximate solution (as in Sec. III)  might be not sufficient.
Note finally, that the RG results for the classical bond percolation problem can be significantly improved by considering a bigger 
cluster\cite{rey80,ber78}. Using the rules, formulated in Sec. II, 
this advanced scheme can be also generalized to quantum percolation.

\acknowledgments
We are grateful to E. V. Tsiper for exptremely valuable advises on computing.

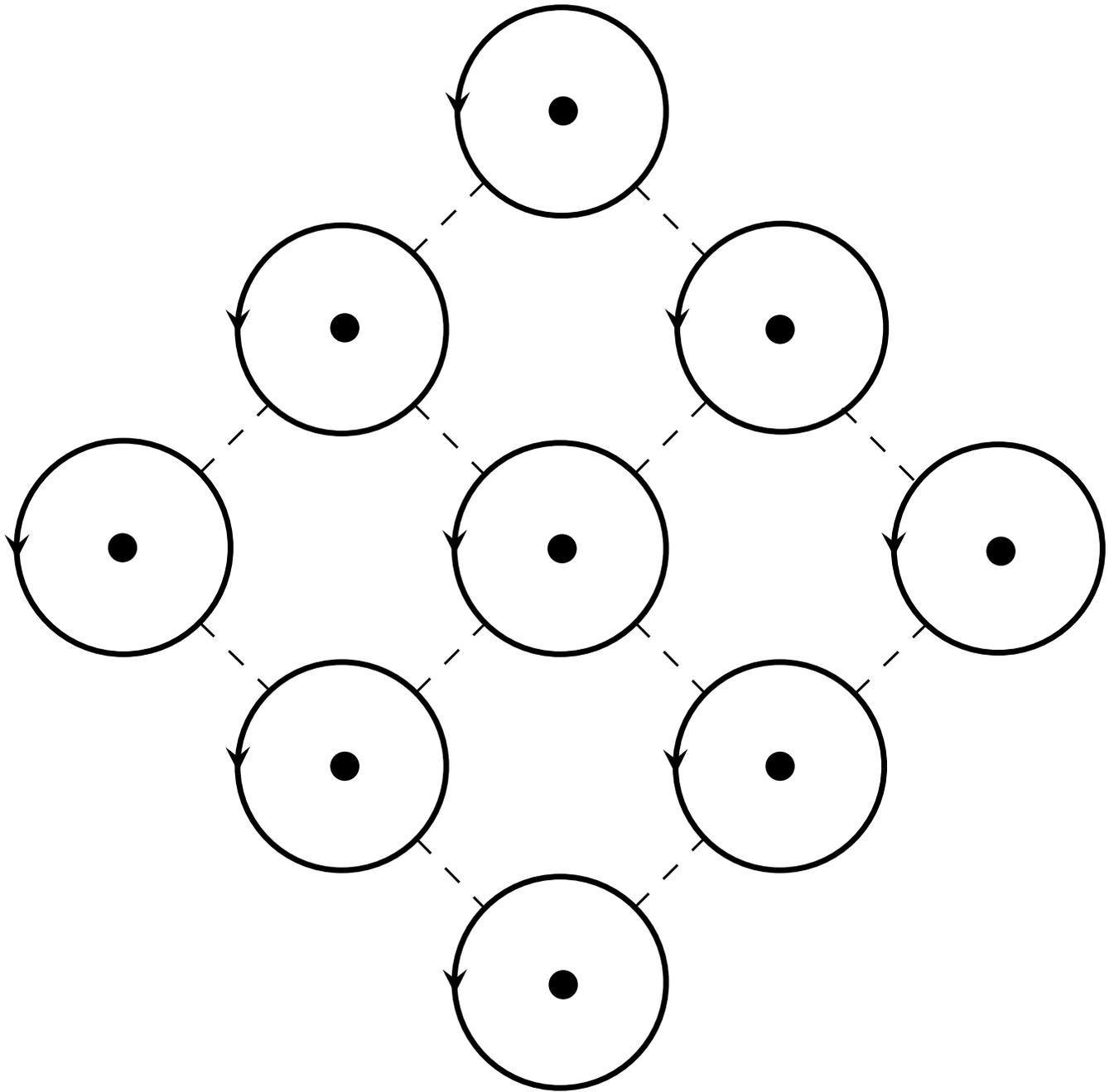
\begin{figure}
\caption{Schematic representation of the network model. Circles show 
equipotentials. Dashed lines show the saddle--points.}
\end{figure}
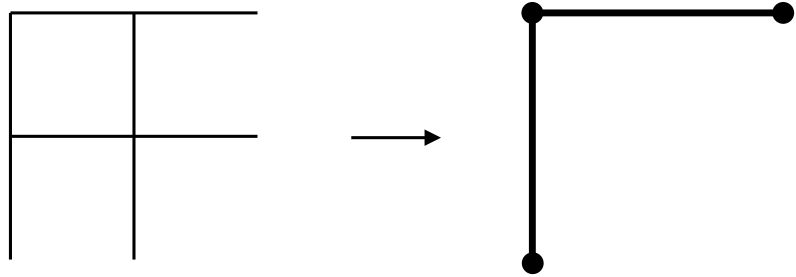
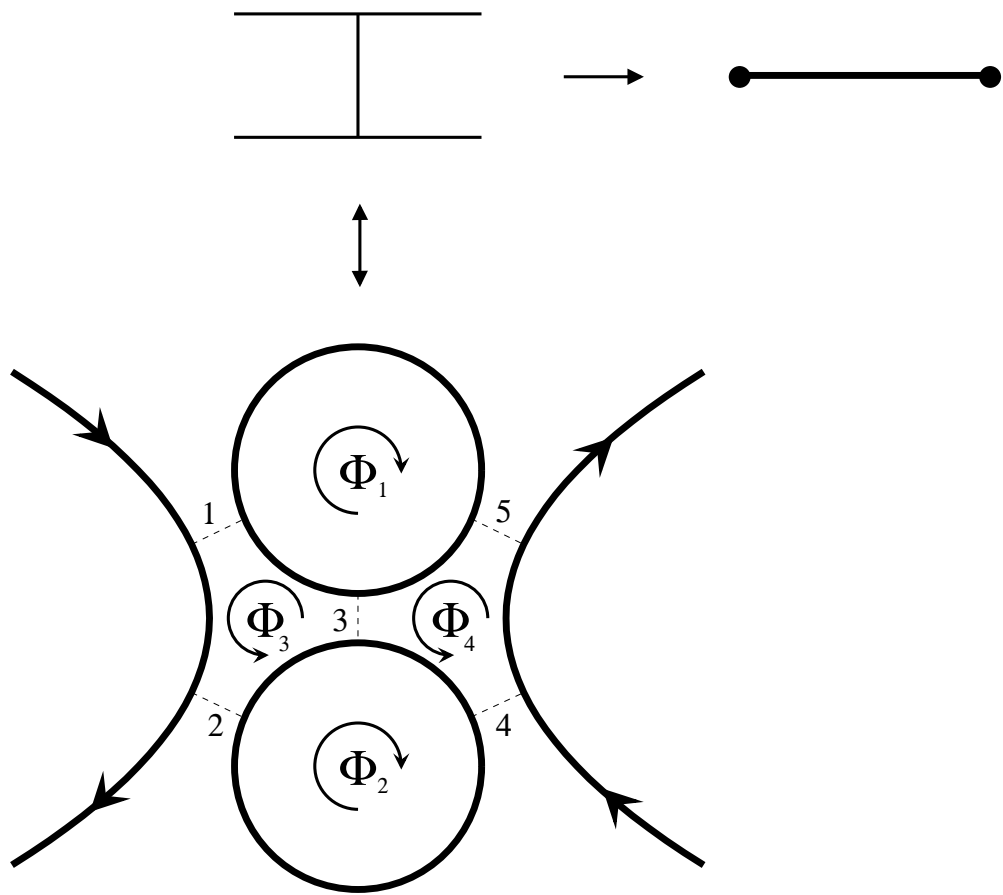
\begin{figure}
\caption{Schematic illustration of the renormalization procedure for the
classical (a) and quantum (b) percolation.}
\end{figure}

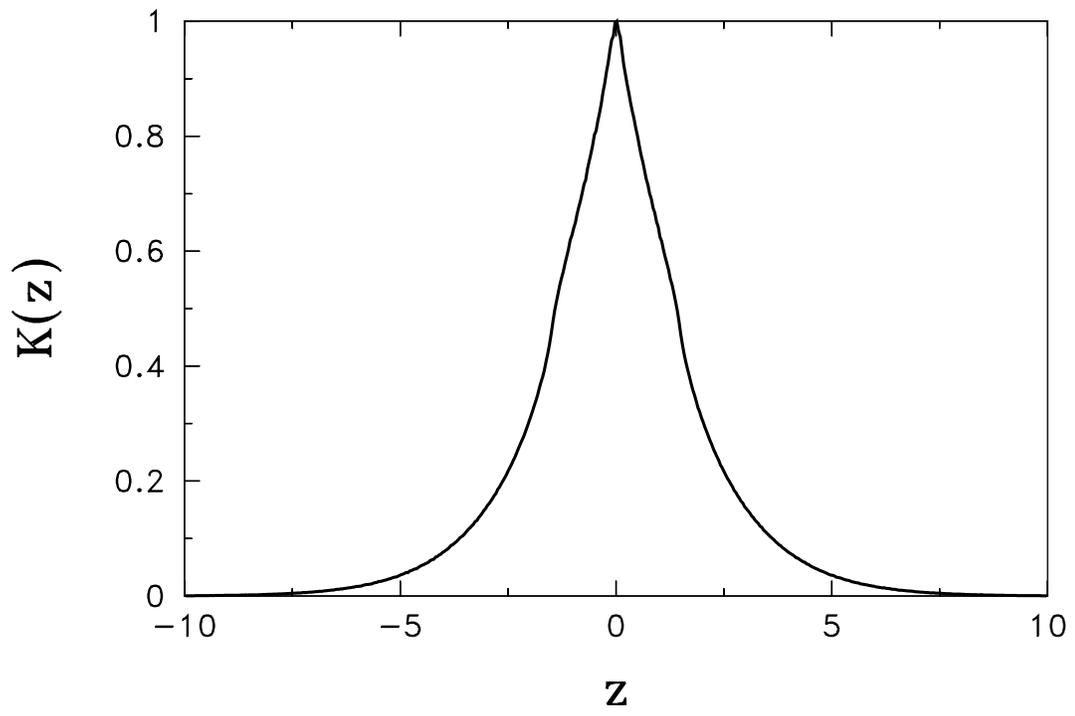
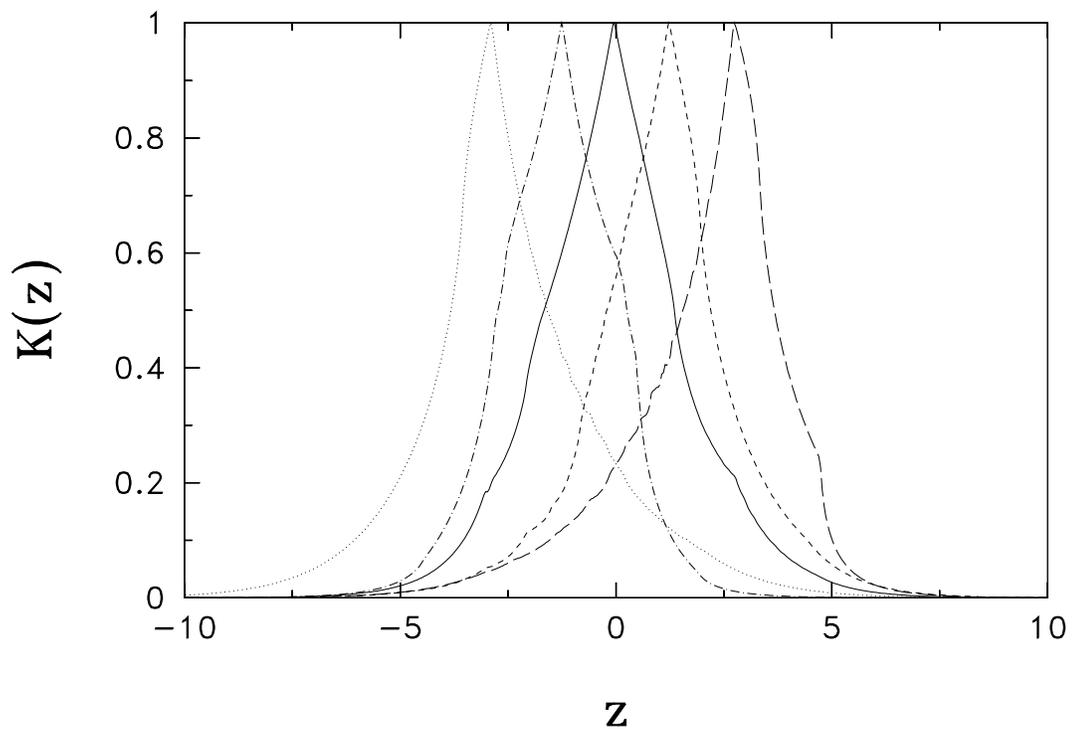
\begin{figure}
\caption{Distribution of heights of the super--saddle--points 
$K(z,\{z_i\})$ is plotted for different sets $\{z_1, z_2, z_3, z_4,
z_5\}$; (a)\{0,0,0,0,0\}; (b)\{-2,2,0,-1,1\} (solid curve), 
\{-1.5,-1.5,-2,-1.5,-1.5\} (dotted curve), \{-1.5,1.5,0,-1.5,-1.5\} (dashed--dotted curve), \{1,-1,2,1,5,0.5\}
(short--dashed curve) \{1,2,2,1,2\} (long--dashed curve).}
\end{figure}

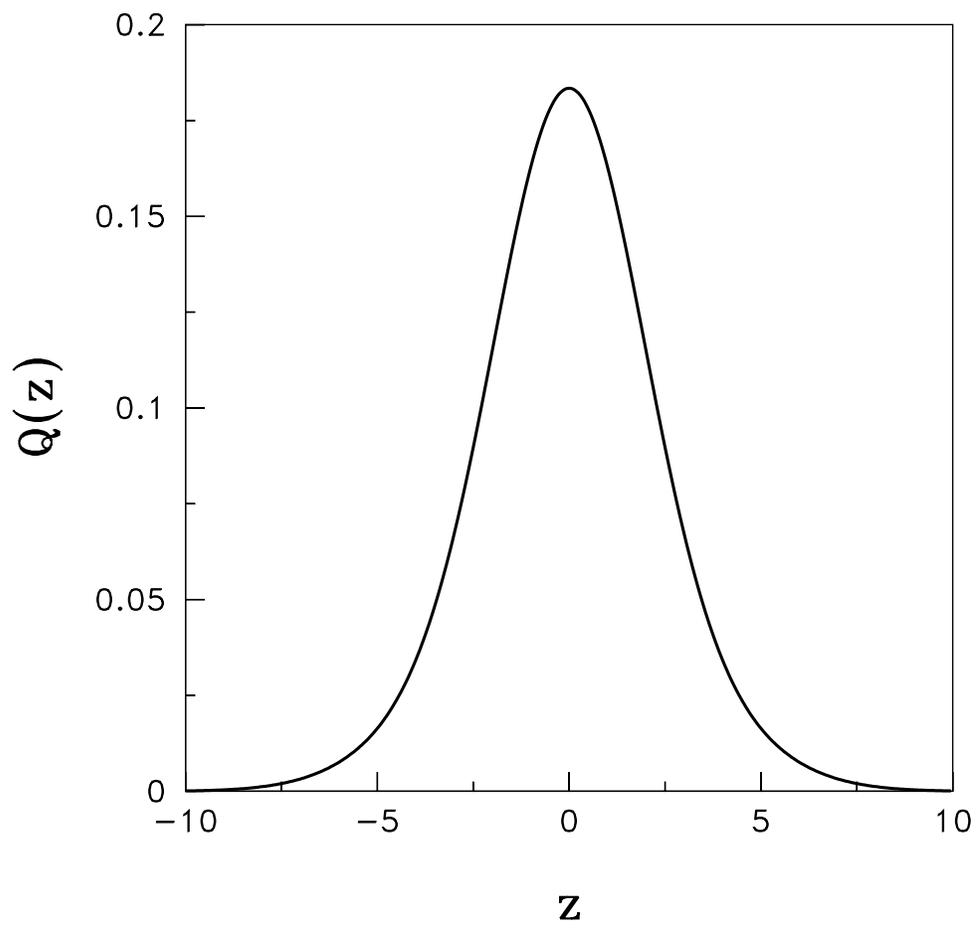
\begin{figure}
\caption{Fixed--point distribution of heights of the super--saddle--points.}
\end{figure}

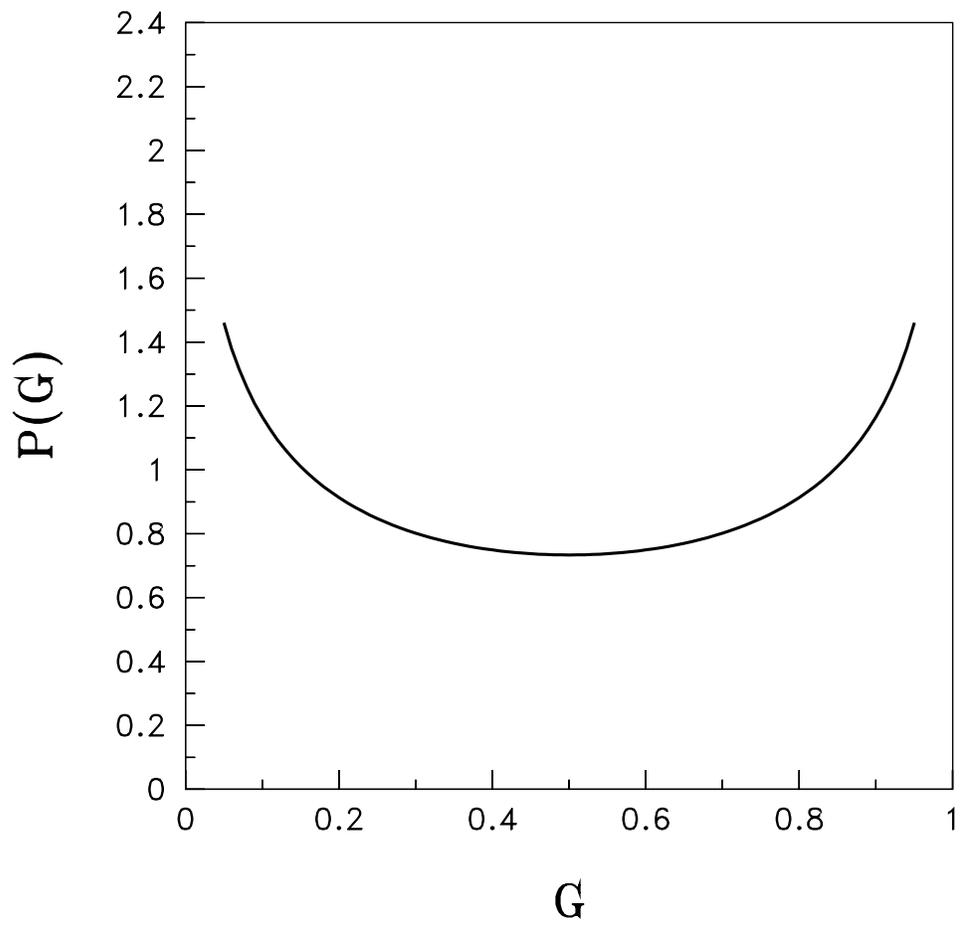
\begin{figure}
\caption{Distribution function of conductance at the quantum Hall transition.}
\end{figure}

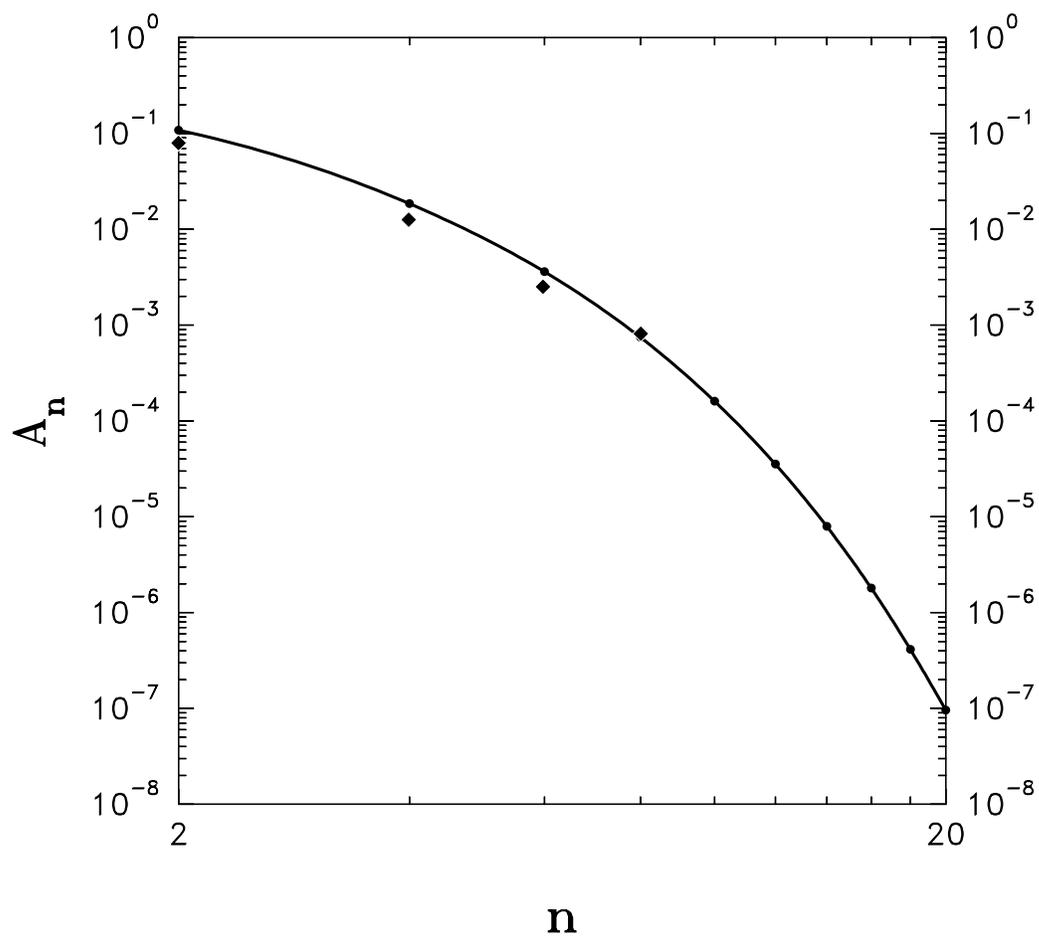
\begin{figure}
\caption{Moments $A_n=\langle (G -\frac{1}{2})^n\rangle$ 
of the conductance distribution
at the transition are shown for different $n$. Diamonds show the
numerical  results of
Ref. 30.}
\end{figure}

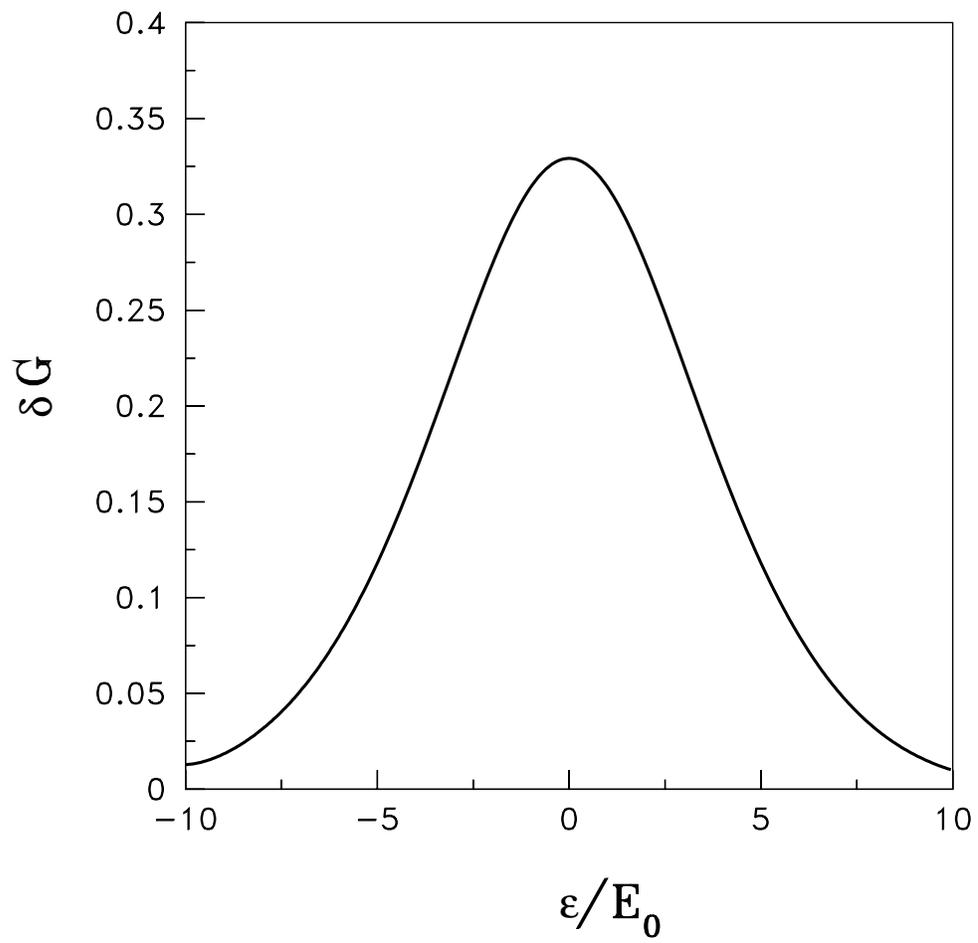
\begin{figure}
\caption{Variance of the conductance fluctuations is plotted v.s. dimensionless energy.}
\end{figure}

\end{document}